\documentclass[aps,prl,twocolumn,showpacs]{revtex4}

\usepackage{graphicx}

\begin{document}

\title{Inherent Relation between Nernst Signal and Nodal Quasiparticle Transport in Pseudogap Region of Underdoped High Temperature Superconductors}

\author{Hai-Hu Wen, Hong Gao, Hao Jin, Lei Shan, Fang Zhou, Jiwu Xiong, Wenxin Ti}

\affiliation{National Laboratory for Superconductivity, Institute
of Physics, Chinese Academy of Sciences, P.~O.~Box 603, Beijing
100080, P.~R.~China}

\date{\today}

\begin{abstract}
In-plane Nernst signal and resistivity have been measured for
three $La_{2-x}Sr_xCuO_4$ single crystals (x=0.09, 0.11 and 0.145)
with the magnetic field parallel to c-axis. A quadratic
temperature dependence of resistivity, i.e., $\rho=\rho_0+aT^2$ is
observed below a certain temperature $T_R$. It is found that the
upper boundary of the Nernst signal $T_n$ coincides with $T_R$,
which points to an inherent relation between the anomalous Nernst
signal and the nodal quasiparticle transport in the pseudogap
region. Finally a phase diagram together with the pseudogap
temperature $T^*$ is presented, which suggests a second energy
scale in the pseudogap region.
\end{abstract}


\maketitle

One of the core issues in high temperature superconductors is the
origin of a pseudogap above $T_c$ in underdoped region. In order
to understand the physics behind the pseudogap, many models have
been proposed, such as resonating valence bond (RVB)\cite{RVB}
theory, spin fluctuation\cite{SpinFluc}, preformed Cooper
pairs\cite{Emery}, charge stripes\cite{EmeryKivelson}, d-density
wave (DDW)\cite{Chakravarty,Affleck}, etc. Among many of them, the
pseudogap state has been considered as a precursor to the
superconducting state. In this precursor state, Cooper pairs have
already formed before the long range phase coherence (or
Bose-Einstein like condensation) is established. Measurements on
the high frequency complex conductivity illustrate that a
short-life phase coherence can persist up to about 30 K above
$T_c$ and these data can be described by the dynamics of thermally
generated topological defects (free vortices)\cite{Corson}. The
Princeton group found that a significant in-plane Nernst
signal\cite{XuZA} appears in the pseudogap region with H$\|$ c.
This may be understood by the phase-slip due to the thermal
drifting of vortex-like excitations. However, it remains unknown
how does this strong Nernst signal relate to the nodal
quasiparticles which are supposed to dominate the in-plane
transport properties in low temperature region. In this Letter we
present the evidence for an inherent relation between the Nernst
signal and nodal quasiparticle transport: The upper boundary
temperature $T_n$ of Nernst signal is found to coincide very well
with a crossover temperature $T_R$ below which a quadratic
temperature dependence of resistivity is observed. Possible
reasons are given to explain this coincidence.

The single crystals measured in this work were prepared by
travelling solvent floating-zone technique. Samples with three
different doping concentrations p=0.09($T_c$=24.4K, as grown,
x=0.09), 0.11($T_c$=29.3K, as grown, x=0.11), 0.145($T_c$=36.1K,
nominal x=0.15) have been investigated. The quality of our samples
has been characterized by x-ray diffraction, and $R(T)$ data
showing a narrow transition $\Delta T_c \leq $ 2 K. For some
samples, the full width at the half maximum (FWHM) of the rocking
curve of (008) peak is only 0.10$^\circ$\cite{ZhouFang}. The
samples have also been checked by AC and DC magnetization and
resistive measurements. Inset(a) of Fig.1 shows a typical curve of
the diamagnetic transition of sample $p=0.11$ measured at $H=20
Oe$.

\begin{figure}
\includegraphics[width=8cm]{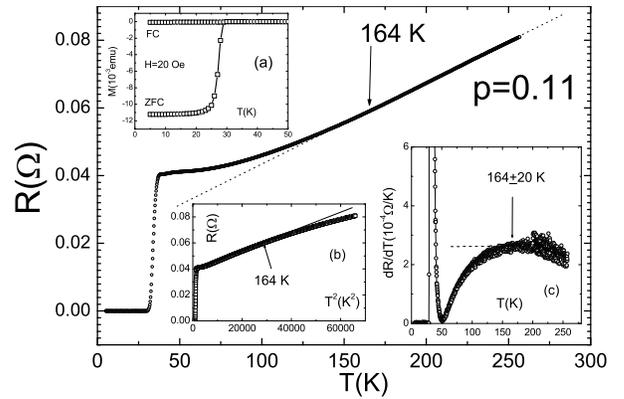}
\caption{Temperature dependence of resistivity for sample $p=0.11$
(shown here as a typical example). An enhancement of resistivity
is observed below a temperature $T_R$ (here about 164 K). Below
$T_R$=164$\pm$20 K a quadratic temperature dependence of
resistivity has been observed (see the linear part of Fig.1(b)).
In Fig.1(c) we show the derivative of resistivity versus $T$. It
is clear that the derivative $dR/dT$ starts to drop down at about
164 K. Similar behavior is found for other two samples with
$p=0.09$ and $p=0.145$ although the $T_R$ are different.}
\label{fig1}
\end{figure}

The resistivity was measured by using standard four-point
technique. For measuring the Nernst effect we adopted the
one-heater-two-thermometer technique. A heating power of 2 mW is
applied to one end of the single crystal and two tiny thermometers
with distance of 1.5 mm are attached to the sample for measuring
the temperatures along the heat flow (longitudinal) direction. The
Nernst voltage is measured through two contacts on two opposite
planes at the symmetric positions. Very small contacting
resistance ($\leq 0.1 \Omega$) has been achieved by using silver
paste. All samples are shaped into a bar structure with dimensions
of $4-5 mm$ (length) $\times$ 1 mm (width) $\times$ 0.5 mm
(thickness). All measurements are based on an Oxford cryogenic
system ( Maglab-12 ) with temperature fluctuation less than
0.04\%$T$ and magnetic fields up to 12 T. During the measurement
for Nernst signal the magnetic field is applied parallel to
$c-axis$ and swept between 7 to -7 T and the Nernst signal $V_N$
is obtained by subtracting the positive field value with the
negative one in order to remove the Faraday signal during sweeping
the field and the possible thermal electric power due to
asymmetric contacts. The Nernst voltage is measured by a Keithley
182-Nanovoltmeter with a resolution of about 5 nV in present case.
In this paper we show the Nernst signal $e_y=E_N/|\bigtriangledown
T|$, where $E_N = V_N/d$ with $d$ the distance between the two
contacts for Nernst voltage, $\bigtriangledown T$ is the
temperature gradient along the heat flow direction.

Fig.1 shows the temperature dependence of resistivity for sample
$p=0.11$ (as a typical example). One can see that a quasi-linear
behavior is observed above a characteristic temperature $T_R$ at
about $164 K$. It is necessary to note that this quasi-linear part
(as marked by the dotted line) is different from the linear part
above $T^*$ which is about $400 K$ for $p=0.11$. Below $T_R$ the
resistivity deviates from the linear behavior. In Fig.1(b) we
present $\rho$ vs. $T^2$. One can see that below about $164\pm20
K$, the curve posses a linear behavior which indicates a relation
of $\rho=\rho_0+aT^2$. In order to determine $T_R$ with higher
accuracy, we present in Fig.1(c) the derivative of resistivity
$dR/dT$ vs. $T$, it is evident that at about $T_R=164\pm20 K$ the
derivative starts to drop down showing a crossover from a
quasi-linear behavior (above $T_R$) to a quadratic behavior below
$T_R$. Similar features are observed for other two samples
($p=0.09$ and $p=0.145$). Note that Ando et al.\cite{AndoT2}
recently reported that the relation $\rho = \rho_0 + aT^2$ in low
temperature region for underdoped La-214 and Y-123 systems, and
proposed that this behavior is characteristic of the nodal
quasiparticles on the so-called Fermi arcs. It is found also by
these authors that $T_R$ increases towards more underdoping.

Fig.2 presents the Nernst signal of sample $p=0.11$ with thermal
stream along [110] direction (the sample was cut with the
longitudinal direction along [110]) and at temperatures from $5 K$
to $220 K$. In order to show the reliability of the measurement,
we repeated the measurements at $20$, $25$ and $220 K$ after a
2-day delay. One can see that the two sets of data for each
temperature coincide very well showing a high reproducibility. In
low temperature region, the Nernst signal is dominated by the
motion of Abrikorsov vortices. One can easily see that the
background when the vortices are freezed (see data at $T=5, 10, 15
K$ in low field region) is actually precisely zero. This is
contrasted by the slight negative background at high temperatures
(above $150 K$). As first discovered by the Princeton
group\cite{XuZA} and later checked by us\cite{WenHHNernst}, a
strong in-plane Nernst signal can be measurable far above $T_c$
(here $T_c$ = 29.3 K for $p=0.11$). Here it shows the same case.
 When $T$ is above $80 K$, the signal
becomes negative and gradually it approaches a background with a
negative slope. And when $T$ is above about $150 K$ the Nernst
signal does not change anymore with $T$, therefore it is
reasonable to define a upper boundary $T_n$ for the Nernst signal
which locates in the region of $150-180 K$ for sample $p=0.11$
with thermal flow direction along [110].

\begin{figure}
\includegraphics[width=8cm]{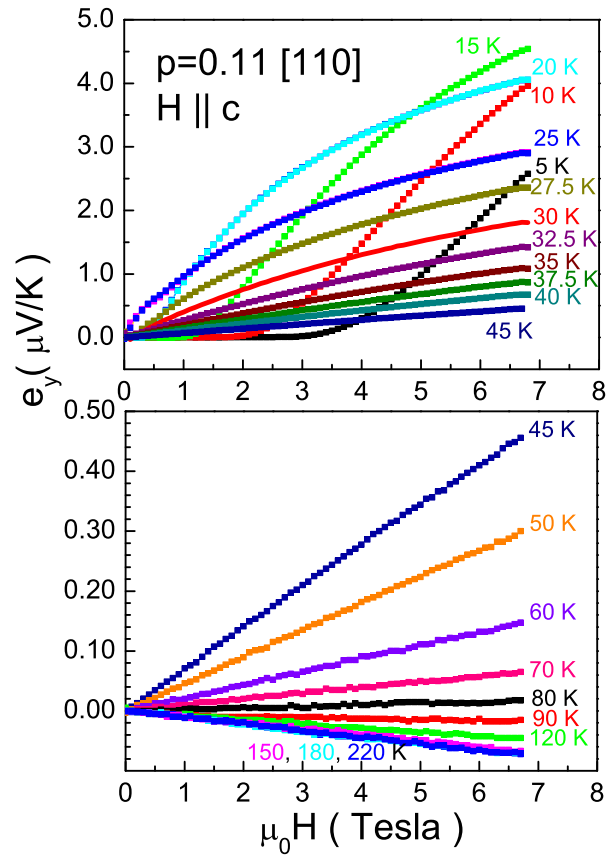}
\caption{ Nernst signal of sample $p=0.11$ at temperatures from
(a) 5 K to 45 K and (b) 45 k to 220 K. The measurement at 20, 25
and 220 K were repeated and data overlap each other for each
temperatures. At temperatures of 5 K, 10 K, 15 K, the vortices are
freezed showing a background with precisely zero resistivity. At
temperatures above about 150 K all data overlap to a background
with a negative slope.} \label{fig2}
\end{figure}

\begin{figure}
\includegraphics[width=8cm]{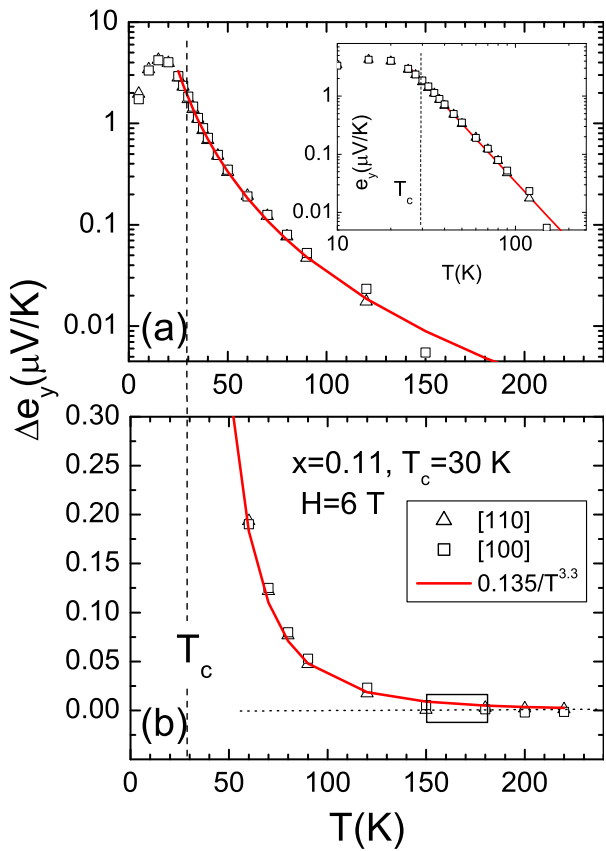}
\caption{ Net Nernst signal $\Delta e_y$ (subtracted with the
background at 220 K) at 6 T shown in (a) semi-logarithmic and (b)
linear scale for samples ($p=0.11$) with thermal stream along
[110] (open triangle) and [100] (open square). From both figures
one can see that the Nernst signal vanishes in the region of 150 K
to 180 K as marked by a rectangular in Fig.3(b). The solid line
shows an empirical relation $\Delta e_y=0.135/T^{3.3}$ which
describes the data above $T_c$ very well but the reason remains
unknown. Inset of Fig.3(a) shows the same data in double
logarithmic scale.} \label{fig3}
\end{figure}

In Fig.3 the Nernst signal from both samples ($p=0.11$ along [110]
and [100]) at 6 T is shown vs. T. The red (solid) curve is an
empirical relation $\Delta e_N=e_N^0/T^{3.3}$ which describes the
data very well up to about $120 K$ and a deviation from this
behavior with a faster dropping rate occurs above about $120 K$.
It is thus safe to conclude that the Nernst signal vanishes in the
region of $150 K$ to $180 K$ as marked by a rectangular in
Fig.3(b). Surprisingly this upper boundary temperature $T_n$ of
Nernst signal coincides rather well with the crossover temperature
$T_R=164\pm 20 K$ as mentioned above. For sample ($p=0.11$) we
have measured on two specimens, one with the heat stream along
[100] and another one along [110]. The results on these two
samples turn out to be almost identical (as shown by open squares
and triangles in Fig.3). Then we checked for other two samples
with doping $p$=0.09,0.145 and found that this coincidence of
$T_R$ and $T_n$ is true also for them. From the recent published
data of Ando et al.\cite{AndoT2} the $T_R$ has also been
determined and shown together with our data in Fig.4. It is clear
that the data from both groups overlap each other. In Fig.4 we
also present the doping dependence of $T_n$ given by the Princeton
group. In the very underdoped region, according to Ando et
al.\cite{AndoT2}, the crossover of resistivity occurs at higher
temperatures. For example, at $p=0.02$, $T_R$ is about $230 K$.
However the upper boundary of Nernst signal is obscured by both
the small Nernst signal and strong thermal power in very
underdoped region. Therefore we don't know whether this
coincidence holds still for very underdoped samples. If taking a
fixed value which is determined by the resolution of the voltmeter
as a criterion for the Nernst signal, according to the recent data
of the Princeton group\cite{WangYYPRB}, the Nernst upper boundary
temperature $T_n$ will drop down in very underdoped region.
Despite the uncertainty in determining the value of $T_n$, in the
doping region of our samples ($p=0.09,0.11,0.145$), the data are
not far from that of the Princeton group.

\begin{figure}
\includegraphics[width=8cm]{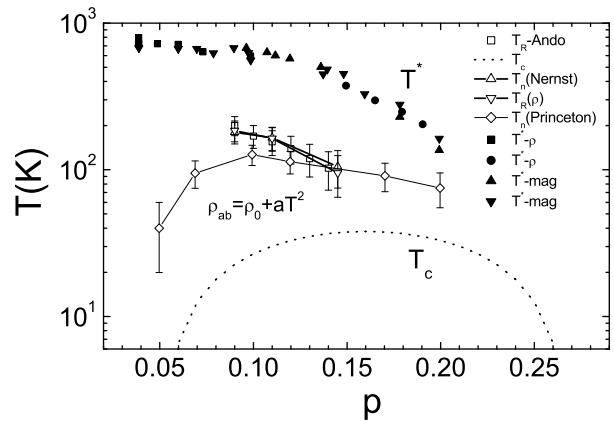}
\caption{ Phase diagram of $La_{2-x}Sr_xCuO_4$ single crystals.
The bottom dotted line represents an empirical curve of
$T_c/38=1-82.6(p-0.16)^2$. The filled symbols represent the
temperatures $T^*$ (quoted from\cite{Tstar}) below which the Fermi
surface is partially gapped. The $T^*-\rho$ was determined from
the resistivity and $T^*-mag$ was determined as the point where
the Knight shift starts to decrease. The open symbols are
determined from our measurement for $T_n$ and $T_R$, $T_R$ from
Ando et al.\cite{AndoT2} and $T_n$ from Wang et
al.\cite{WangYYPRB}. The upper boundary of Nernst signal ($T_n$)
and the crossover temperature of resistivity $T_R$ coincide rather
well in the doping regime of our present samples. } \label{fig4}
\end{figure}

In the following we will try to understand the coincidence of
$T_n$ and $T_R$ in the intermediate doping region based on several
possible pictures. We first discuss the scenario of vortices. It
is known that the in-plane resistivity normally reflects the
dynamics of the nodal quasiparticles, while the Nernst signal is
induced by the thermal drifting of vortices as suggested by the
Princeton group, therefore it seems difficult to relate one and
another. However one way to understand this point is that the
nodal quasiparticles partially form Cooper pairs and induce a
precursor superconducting state which has strong phase fluctuation
with vortices as one of the important excitations. Therefore below
this specific temperature the Nernst signal which senses the
vortex motion starts to appear and the temperature dependence of
resistivity changes behavior (due to probably the reduction of the
effective charge carriers). As a partial support to this picture,
it was shown that this strong in-plane Nernst signal may be
explained by the superconducting fluctuation which has the
Kosterlitz-Thouless vortex-anti-vortex pairs as the topological
excitations\cite{Ussishkin}. The second picture close to this
vortex scenario is based on the bosonic RVB\cite{WengZY} which
predicts the co-existence of spinon vortices and quasiparticles
below $T_n$ and the Nernst signal is induced by the motion of
these spinon vortices. This picture naturally predicts a second
energy scale ($T_n$ or $T_R$) and different type of vortices below
and above $T_c$ (still lacking of experimental evidence for the
spinon vortices). However both pictures are facing a common
difficulty: No any trace of vortex (or spinon vortex) flow
resistivity has been reported far above $T_c$. One explanation to
this point would be that the resistivity induced by quasiparticle
scattering is overwhelmingly larger than the flux flow
resistivity. Meanwhile it remains to be understood why the
temperature dependence of resistivity should be precisely
quadratic. Two close relatives of the picture based on vortices
are the one with a mixture of composite charge
carriers\cite{Kim,Levin} or the pair density wave
(PDW)\cite{ZhangSCPDW} (coexistence of Cooper pairs and free
electrons below $T_n$). The paired electrons tend to localize and
the rest free electrons contribute to the electrical conductivity.
According to Ando et al.\cite{AndoT2}, the relation
$\rho=\rho_0+aT^2$ may not be simply due to the scattering rate
$1/\tau \propto T^2$ of the nodal quasipartcles since the
effective charge carrier density changes with temperature even
below $T_R$. A naive explanation for $\rho=\rho_0+aT^2$ in this
region would be that the nodal quasiparticle density (roughly
proportional to the length of Fermi arc\cite{Norman} or the
electronic specific heat coefficient $\gamma$\cite{Loram}) drops
down linearly with $T$, but the scattering rate $1/\tau$ of nodal
quasiparticles with features close to Dirac fermions is
proportional to $T^3$. One more possibility for the coincidence of
$T_R$ and $T_n$ relies on a exotic picture that some kind of
unconventional charge density wave (UCDW)\cite{Maki} occurs at
$T_n$. Below $T_n$ or $T_R$ the quasiparticle spectrum on top of
this UCDW will contribute a strong Nernst signal, while the nodal
quasiparticles are responsible for the in-plane resistivity, thus
both the in-plane Nernst signal and resistivity will certainly
correlate each other. This picture needs both theoretical
justification and experimental evidence, especially a recent
calculation seems ruling out the D-density wave as one of the
possible causes\cite{DDWUssishkin}. Since the Nernst signal is
detectable only below $T_R$, it may have no direct relation with
the "flat band" near $(\pi,0)$ as observed by ARPES since it
influences the transport properties only at high
temperatures\cite{Flatband}. The coincidence of $T_R$ and $T_n$
found in our experiment strongly suggests an inherent relation
between the Nernst signal and the nodal quasiparticle transport.
This may also imply a second energy scale in the pseudogap region.

In conclusion, the in-plane Nernst signal and a quadratic
temperature dependence of resistivity are found to occur in the
same temperature region above $T_c$ in the intermediate underdoped
regime. This suggests a close and inherent relation between Nernst
signal and the nodal quasiparticle transport. Possible reasons are
given to explain this coincidence. This coincidence may be
obscured by other effects in other doping regimes. Our observation
may imply a second energy scale or temperature within the
pseudogap region.

This work is supported by the National Science Foundation of China
, the Ministry of Science and Technology of China, the Chinese
Academy of Sciences. We thank Dr. Yoichi Ando and his group
(CRIEP, Komae, Tokyo, Japan) for providing us the sample with
$p$=0.145. We are grateful to Z. X. Zhao, Z. Y. Weng, N. Nagaosa,
T. Xiang and Z. A. Xu for fruitful discussions.


Correspondence should be addressed to hhwen@aphy.iphy.ac.cn

\end{document}